\documentclass[3p,times,twocolumn]{elsarticle}

\usepackage{ecrc}

\volume{00}

\firstpage{1}

\journalname{Nuclear Physics B Proceedings Supplement}

\runauth{F. Mahmoudi, A. Arbey}

\jid{nuphbp}

\jnltitlelogo{Nuclear Physics B Proceedings Supplement}

\usepackage{amssymb}

\biboptions{sort&compress}

\usepackage[figuresright]{rotating}

\begin{document}

\begin{frontmatter}




\title{Complementarity of direct and indirect searches in the pMSSM\tnoteref{label1}}  
\tnotetext[label1]{Based on the talk by F.M. at the Fifth Capri Workshop on the interplay of flavour physics with electroweak symmetry breaking and dark matter, Capri, May 2014.}


\author{Farvah Mahmoudi}
\ead{nazila@cern.ch}
\author{Alexandre Arbey}
\ead{alexandre.arbey@ens-lyon.fr}
\address{Universit{\' e} de Lyon, Universit{\' e} Lyon 1, F-69622 Villeurbanne Cedex, France;\\
Centre de Recherche Astrophysique de Lyon, Saint Genis Laval Cedex, F-69561, France;\\ CNRS, UMR 5574;
Ecole Normale Sup{\' e}rieure de Lyon, France}
\address{CERN Theory Division, Physics Department, CH-1211 Geneva 23, Switzerland}

\begin{abstract}
We explore the pMSSM parameter space in view of the constraints from SUSY and monojet searches at the LHC, from Higgs data and flavour physics observables, as well as from dark matter searches. We show that whilst the simplest SUSY scenarios are already ruled out, there are still many possibilities left over in the pMSSM. We discuss the complementarity between different searches and consistency checks which are essential in probing the pMSSM and will be even more important in the near future with the next round of data becoming available.
\end{abstract}


\end{frontmatter}




\section{Introduction}

\begin{table}[h!]
\begin{center}
\begin{tabular}{|c|c|}
\hline
~~~~Parameter~~~~ & ~~~~Range (in GeV)~~~~ \\
\hline\hline
$\tan\beta$ & [1, 60]\\
\hline
$M_A$ & [50, 2000] \\
\hline
$M_1$ & [-3000, 3000] \\
\hline
$M_2$ & [-3000, 3000] \\
\hline
$M_3$ & [50, 3000] \\
\hline
$A_d=A_s=A_b$ & [-10000, 10000] \\
\hline
$A_u=A_c=A_t$ & [-10000, 10000] \\
\hline
$A_e=A_\mu=A_\tau$ & [-10000, 10000] \\
\hline
$\mu$ & [-3000, 3000] \\
\hline
$M_{\tilde{e}_L}=M_{\tilde{\mu}_L}$ & [0, 3000] \\
\hline
$M_{\tilde{e}_R}=M_{\tilde{\mu}_R}$ & [0, 3000] \\
\hline
$M_{\tilde{\tau}_L}$ & [0, 3000] \\
\hline
$M_{\tilde{\tau}_R}$ & [0, 3000] \\
\hline
$M_{\tilde{q}_{1L}}=M_{\tilde{q}_{2L}}$ & [0, 3000] \\
\hline
$M_{\tilde{q}_{3L}}$ & [0, 3000] \\
\hline
$M_{\tilde{u}_R}=M_{\tilde{c}_R}$ & [0, 3000] \\
\hline
$M_{\tilde{t}_R}$ & [0, 3000] \\
\hline
$M_{\tilde{d}_R}=M_{\tilde{s}_R}$ & [0, 3000] \\
\hline
$M_{\tilde{b}_R}$ & [0, 3000] \\
\hline
\end{tabular}
\caption{pMSSM scan ranges.\label{tab:pmssm}}
\end{center}
\end{table}%

\begin{figure*}[t!]
\begin{center}
\includegraphics[width=7.5cm]{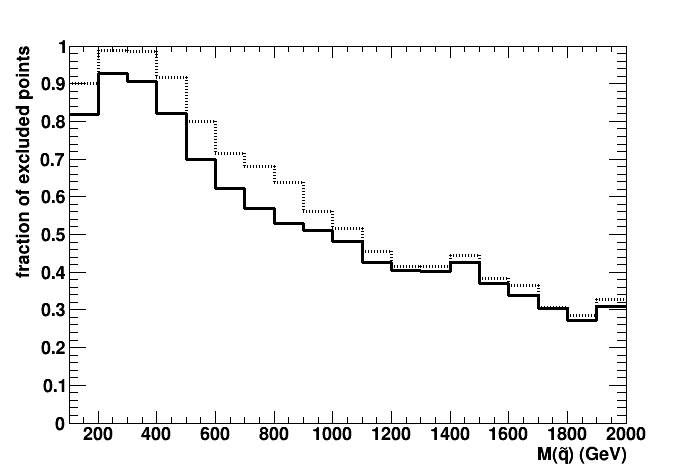}\quad
\includegraphics[width=7.5cm]{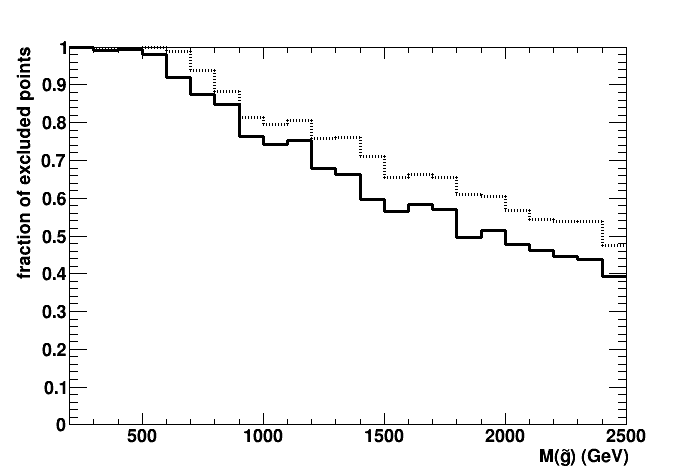}
\caption{Fraction of pMSSM points excluded by the LHC SUSY searches (solid line) and complemented by the monojet searches (dotted line) as a function of the lightest squark mass (left panel) and of the gluino mass (right panel).\label{fig:LHC_searches}}
\end{center}
\end{figure*}%

Data from the first run of the LHC and from dark matter (DM) direct detection experiments provide us already with important constraints on the supersymmetric parameter space. We consider the phenomenological MSSM (pMSSM), which is the most general MSSM set-up with R-parity and CP conservation respecting the minimal flavour violation (MFV) paradigm. With its 19 independent parameters, the pMSSM offers sufficient freedom to the masses and couplings to explore the supersymmetric
parameter space and the implications of the recent particle and astroparticle physics data in a largely unbiased way. The 19 parameters are scanned over in the ranges given in Table~\ref{tab:pmssm} following the methodology described in Refs.~\cite{Arbey:2011un,Arbey:2011aa}.

The SUSY mass spectra and decay branching fractions are computed using SOFTSUSY \cite{Allanach:2001kg}, HDECAY \cite{Djouadi:1997yw} and SDECAY \cite{Muhlleitner:2003vg}. The Higgs boson production rates are calculated with HIGLU \cite{Spira:1996if}, bbh@nnlo \cite{Harlander:2003ai} and HQQ \cite{Spira:1997dg}. Flavour physics observables are computed using SuperIso \cite{Mahmoudi:2007vz} and the neutralino relic density with SuperIso Relic \cite{Arbey:2009gu}. Dark matter direct detection observables are evaluated with micrOMEGAs~\cite{Belanger:2008sj}. The constraints from the LHC monojet and SUSY searches are assessed using MadGraph~\cite{Alwall:2011uj}, PYTHIA~\cite{Sjostrand:2007gs} and Delphes~\cite{deFavereau:2013fsa}.

In the following, for the relic density, we apply the upper bound of the cold dark matter density measured by Planck \cite{Ade:2013zuv}, allowing for the possibility that dark matter is composed of multiple components or that the early Universe properties differ from that of the standard cosmological model \cite{Arbey:2008kv,Arbey:2009gt}. The constraints from flavour physics and low energy data are also applied as described in~\cite{Arbey:2013aba}.

\section{SUSY and monojet searches at the LHC}

The direct searches for MET signatures with jets and leptons provide important constraints on the strongly interacting SUSY particle masses. However, these searches 
lose sensitivity in the regions where the mass splittings are small. Such cases often lead to soft jets which are invisible in the detectors. Monojet searches on the other hand are efficient in such situations and can be used to improve the sensitivity~\cite{Arbey:2013iza}. 
In Fig.~\ref{fig:LHC_searches}, the effect of the SUSY (solid line) and monojet (dotted line) search results on the squark and gluino masses is presented. It is remarkable that while the limits on the squark and gluino masses are above 1.5--2 TeV in the constrained or simplified MSSM scenarios, in a more general scenario such as the pMSSM with no ad hoc universality assumption for the SUSY particle masses, this is not any more the case. The figure reveals that many model points with masses below 1 TeV are still allowed and that 30\% of the points have still squark masses below 800 GeV and gluino masses below 1.5 TeV.

\begin{figure*}[t!]
\begin{center}
\includegraphics[width=7.cm]{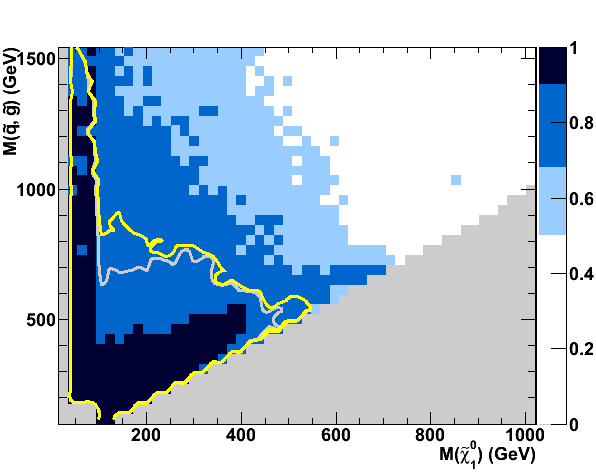}\quad\quad\quad
\includegraphics[width=7.cm]{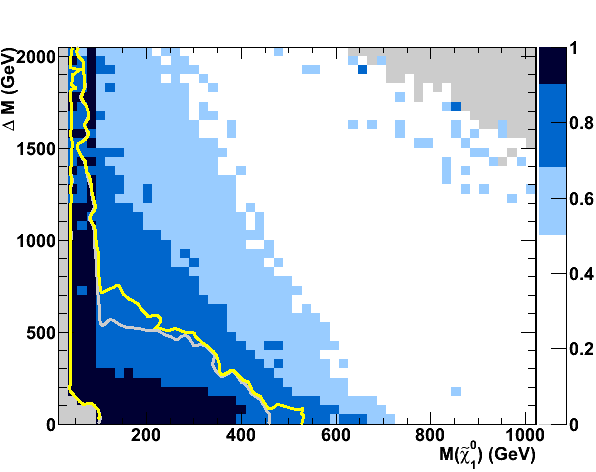}
\caption{Fraction of pMSSM points excluded by the combination of the LHC jets/leptons+MET, monojet analyses and direct DM 
searches in the ($M_{\tilde{\chi}^0_1}$ , $M_{\tilde{q},\tilde{g}}$) (left panel) and ($M_{\tilde{\chi}^0_1}$ , $\Delta M=M_{\tilde{q},\tilde{g}}-M_{\tilde{\chi}^0_1}$) (right panel) planes, where $M_{\tilde{q},\tilde{g}}$ is the lightest of the gluino and squark masses. The lines give the parameter region where 
68\% of the pMSSM points are excluded by the SUSY searches alone (grey line) and the combination with monojet searches (yellow line) (from Ref.~\cite{Arbey:2013iza}).
\label{fig:LHC_searches2D}}
\end{center}
\end{figure*}

To investigate further the complementarity between the SUSY and monojet searches we demonstrate in Fig.~\ref{fig:LHC_searches2D} the fraction of excluded points, for the lightest squark/gluino mass vs. neutralino mass, and for the mass splitting between the lightest squark/gluino and neutralino vs. the neutralino mass. The plots show that squark and gluino masses below 400 GeV are now strongly disfavoured regardless of the lightest neutralino mass, and that the monojet searches are very sensitive in the region of small mass splittings and can improve the reach by more than 100 GeV.

\section{Higgs searches}

\begin{figure}[t!]
\begin{center}
\includegraphics[width=7.5cm]{{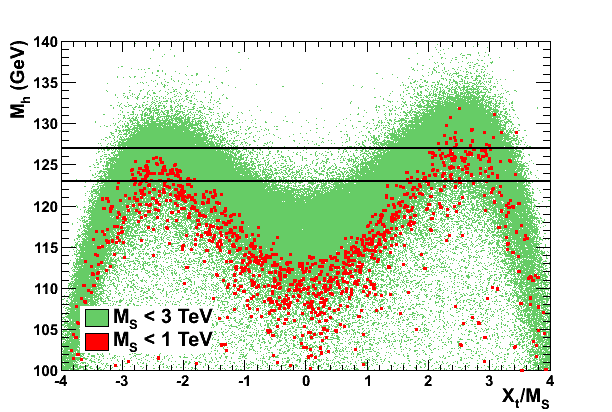}}
\caption{Lightest CP-even Higgs mass in the pMSSM as a function of the stop mixing parameter $X_t$ for $M_S$ smaller than 1 TeV (red points) and 3 TeV (green points) (from Ref.~\cite{Arbey:2011ab}).\label{fig:mh_pmssm}}
\end{center}
\end{figure}

The discovery of a Higgs boson of $\sim\!\!126$ GeV at the LHC has strong implications for supersymmetry. In particular, it is generally assumed that the discovered Higgs boson corresponds to the lightest MSSM CP-even state. 
The mass of the lightest CP-even Higgs boson is related in the MSSM to the stop sector parameters, and in particular to the geometric average of the stop masses $M_S$ and to the stop mixing parameter $X_t$~\cite{Djouadi:2005gj}. In Fig.~\ref{fig:mh_pmssm}, we show that for $M_S < 1$ TeV, a Higgs mass of 126 GeV can only be reached for $M_S \!\sim\! \sqrt6 \,|X_t|$, corresponding to the maximal mixing scenario, while other mixings can be allowed for larger $M_S$ \cite{Arbey:2011ab}.

\begin{figure}[t!]
\begin{center}
\includegraphics[width=7.5cm]{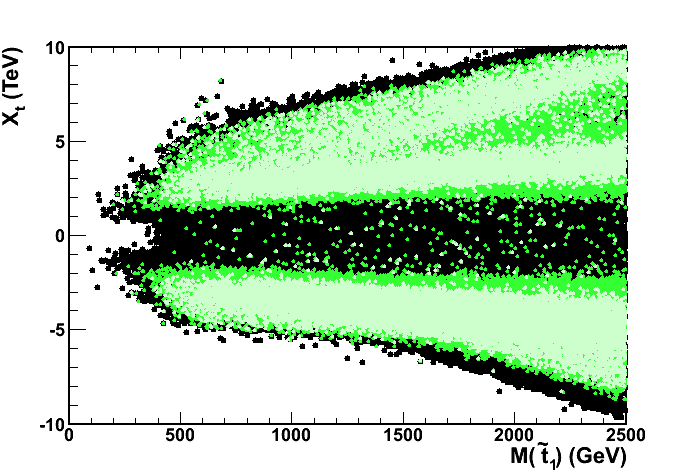}
\caption{Distribution of the pMSSM points in the plane ($M_{\tilde{t}_1},X_t$). The black dots show all the accepted points, those in dark (light) green the points compatible with the Higgs mass and rate constraints at 90\% (68\%) C.L. (from Ref.~\cite{Arbey:2012bp}).
\label{fig:XtMt}}
\end{center}
\end{figure}

In Fig.~\ref{fig:XtMt} we present the distribution of the pMSSM points compatible with the $h$ boson mass and the observed yields, in the ($M_{\tilde{t_1}},X_t$) plane~\cite{Arbey:2012bp,Arbey:2012dq}. We notice again that small values of mixing parameter $|X_t|$ are clearly disfavoured, and that stop masses as low as 400 GeV are still compatible with the data. 

\begin{figure}[t!]
\begin{center}
\includegraphics[width=7.5cm]{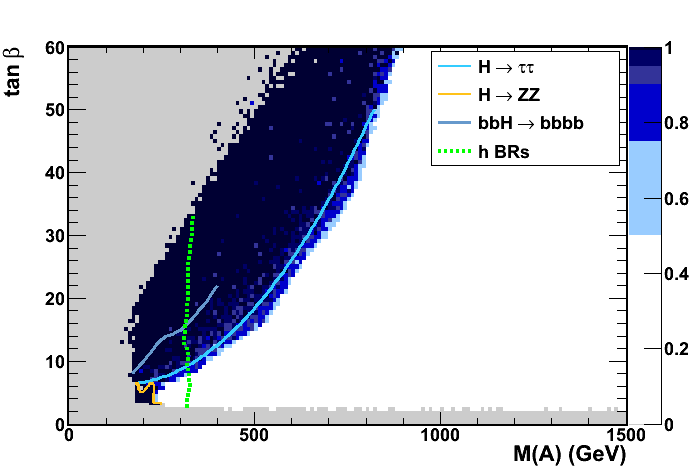}
\caption{Constraints on the $(M_A,\tan\beta)$ parameter plane from the $H$ to $\tau \tau$ and $ZZ$ 
channels for the 8~TeV data set. The colour scale gives the fraction of pMSSM points excluded at each $M_A$ and $\tan \beta$ value. The contours show the limits corresponding to 95\% or more of the points excluded. The 90\% C.L. constraint from the Higgs signal strengths is also shown (green dotted line). The grey region has no pMSSM points passing the $B_s \rightarrow \mu^+ \mu^-$, direct DM searches and $M_h$ constraints (from Ref.~\cite{Arbey:2013jla}).\label{fig:hBR}}
\end{center}
\end{figure}

The determination of the Higgs branching fractions to two photons, two $W$ or $Z$ bosons, and to two fermions has strong implications, in particular when complemented by the heavier Higgs state searches \cite{Arbey:2013jla}. In Fig.~\ref{fig:hBR}, we show the constraints on the $(M_A,\tan\beta)$ plane from the determination of the light Higgs branching ratios and from searches for heavier states. As can be seen, the combination of the $H/A \to \tau^+ \tau^-$ channel and the mass and measured branching ratios for the lightest $h$ boson exclude the region with $M_A < 320$ GeV for all values of $\tan\beta$. The $ZZ$ channel, and to a lesser extent the $WW$ channel, close the low $M_A$ corner from $\tan\beta \simeq 2$ up to the $\tau \tau$ limit for $M_A < 230$ GeV. 

The constraints derived by the study of the Higgs sector are becoming an essential part of the probe of the SUSY parameter space at the LHC.

\section{Flavour physics}

In addition to the direct searches, indirect information from the flavour physics observables are of great interest in probing the supersymmetric parameter space. 
One of the most constraining observables is the rare leptonic decay of $B_s \to \mu^+ \mu^-$, as it can receive supersymmetric contributions scaling with $\tan^6\beta/M^4_A$, increasing the branching ratio by orders of magnitude. This branching ratio has been measured by LHCb and CMS \cite{CMSandLHCbCollaborations:2013pla} with the central value well compatible with the Standard Model prediction, strongly limiting large deviations due to SUSY contributions \cite{Arbey:2012ax}. 

\begin{figure}[t!]
\begin{center}
\includegraphics[width=7.5cm]{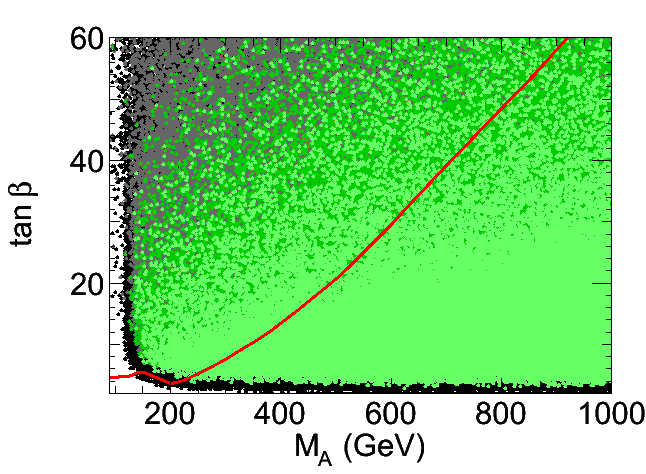}
\includegraphics[width=7.5cm]{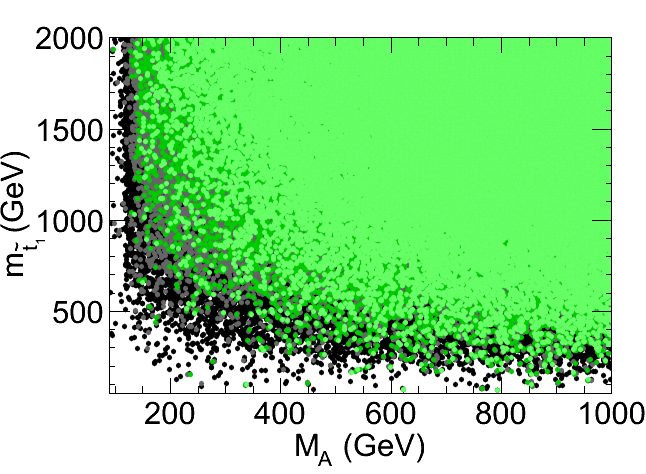}
\caption{Constraints from BR($B_s \to \mu^+ \mu^-$) in the ($M_A,\tan\beta$) (upper panel) and ($M_A,M_{\tilde{t}_1}$) (lower panel) parameter planes. The black points corresponds to all the valid pMSSM points and those in grey to the points for which $123 < M_h < 129$ GeV. The dark green points in addition are in agreement with the latest BR($B_s \to \mu^+ \mu^-$) range, while the light green points are in agreement with the prospective ultimate LHCb BR($B_s \to \mu^+ \mu^-$) range. The red line indicates the region excluded at 95\% C.L. by the CMS $H/A\to\tau^+\tau^-$ searches~\cite{CMS-HIG-2012-050} (from Ref.~\cite{Arbey:2012ax}).\label{fig:Bsmumu}}
\end{center}
\end{figure}

The impact of the present and future determinations of BR($B_s \to \mu^+ \mu^-$) on the parameters most sensitive to its rate, ($M_A,\tan\beta$) and ($M_A,M_{\tilde{t}_1}$), is shown in Fig.~\ref{fig:Bsmumu}, where we give all the valid pMSSM points from our scan (black points), those with $123 < M_h < 129$ GeV (grey points) and, highlighted in green, those in agreement with the present BR($B_s \to \mu^+ \mu^-$) range and the estimated ultimate constraint at 95\% C.L.~\cite{Arbey:2012ax}.
The constraints from BR($B_s \to \mu^+ \mu^-$) affect the same pMSSM region, at large values of $\tan\beta$ and small values of $M_A$, also probed by the $H/A \to \tau^+ \tau^-$ direct Higgs searches. As demonstrated in the previous section, the search for the $H/A \to \tau^+ \tau^-$ decay has already excluded a significant portion of the parameter space where large effects on BR($B_s \to \mu^+ \mu^-$) are expected. 
We also note that the stop sector is further constrained by direct searches in $b$-jets + MET channels, which disfavour small values of $M_{\tilde{t}_1}$. The figure shows that it is difficult for $M_A$ and $M_{\tilde{t}_1}$ to be simultaneously light.

%

\section{Dark matter direct detection}

Indirect constraints can also be obtained from dark matter direct detection experiments. They rely on the measurement of the recoil energy of a nucleus with which a dark matter particle may interact. The interpretation of dark matter direct search results in terms of scattering cross sections of dark matter with matter is however subject to strong assumptions on the local density and velocity of dark matter. Other dark matter observables can be considered additionally. The dark matter relic density is sensitive to the annihilation cross section of neutralinos, but also of other SUSY particles which are close in mass to the neutralino. 
Other constraints can come from indirect detection, which relies on the detection of SM particles produced through the annihilation of dark matter particles. The results suffer strongly from astrophysical assumptions and therefore we do not consider them here.

A well motivated candidate in the MSSM for dark matter is the lightest neutralino. Its scattering with matter is expected to be mediated mainly by a $Z$ boson or a neutral Higgs boson. For this reason, direct detection also probes the $(M_A,\tan\beta)$ plane which is also constrained by heavy Higgs searches and flavour physics observables as discussed in the previous sections. 
This redundancy obtained through searches in multiple channels sensitive in the same regions is essential to fully probe the ($M_A,\tan\beta$) parameter space as it is possible in specific cases to furtively evade one of the constraints, and also for the purpose of consistency checks.

The LUX collaboration presented recently stringent limits on the scattering cross sections of dark matter particles with matter \cite{Akerib:2013tjd}. On the other hand, several direct detection experiments had reported excess of events in the light dark matter regions of a few GeV up to tenths of GeV~\cite{Bernabei:2010mq,Aalseth:2011wp,Angloher:2011uu,Ahmed:2010wy}. While it is possible to explain such excesses in the pMSSM \cite{Arbey:2012na,Arbey:2013aba}, the LUX limits seem to exclude the regions reported by those experiments. 

\begin{figure}[t!]
\begin{center}
\includegraphics[width=7.cm]{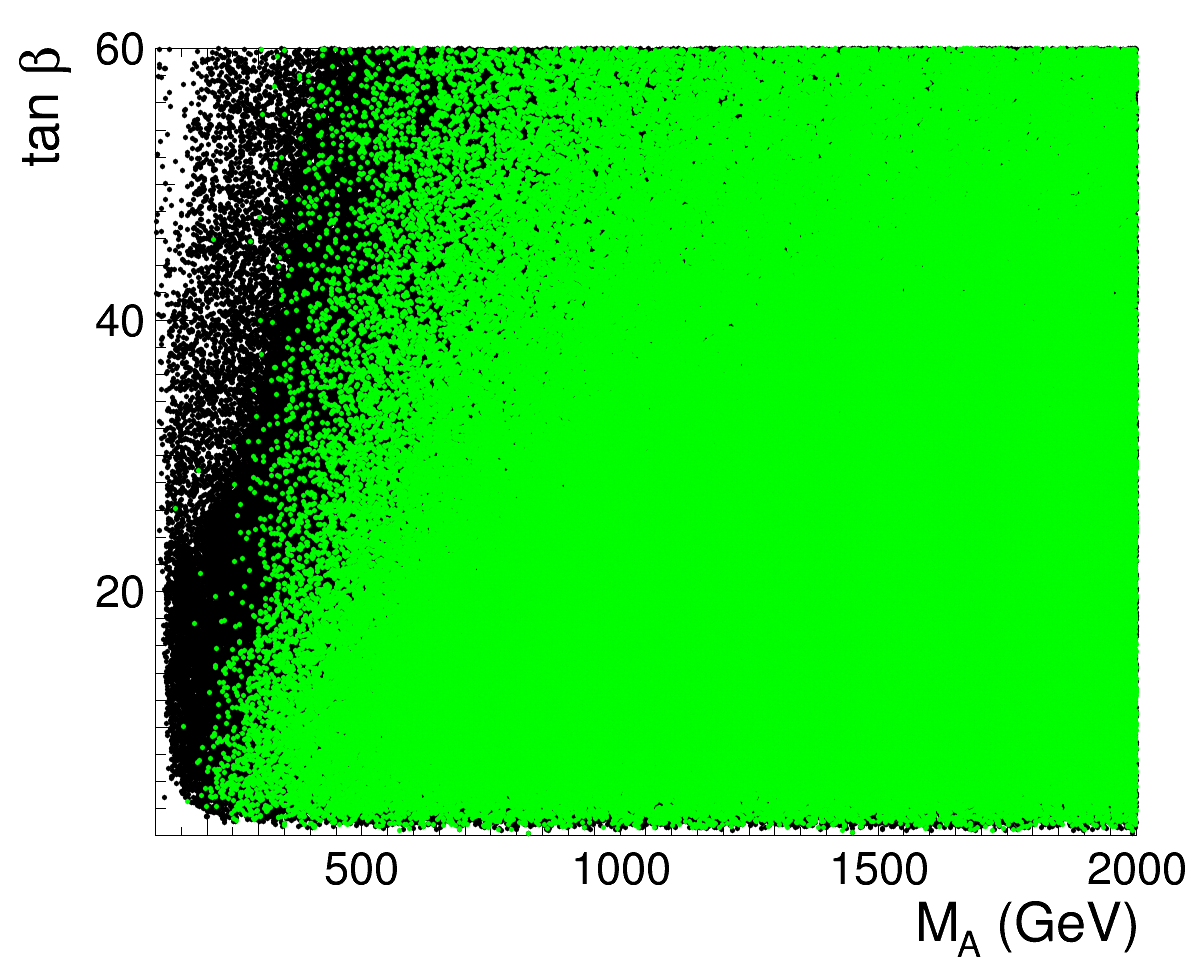}

\caption{Effects on the $(M_A,\tan\beta)$ parameter plane of the constraints from LUX on the dark matter direct detection spin-independent scattering cross section of a neutralino with a proton. The green points corresponds to all the valid pMSSM points and those in black to the points excluded by the LUX limits.\label{fig:LUX}}
\end{center}%
\end{figure}

\begin{figure}[t!]
\begin{center}
\includegraphics[width=7.cm]{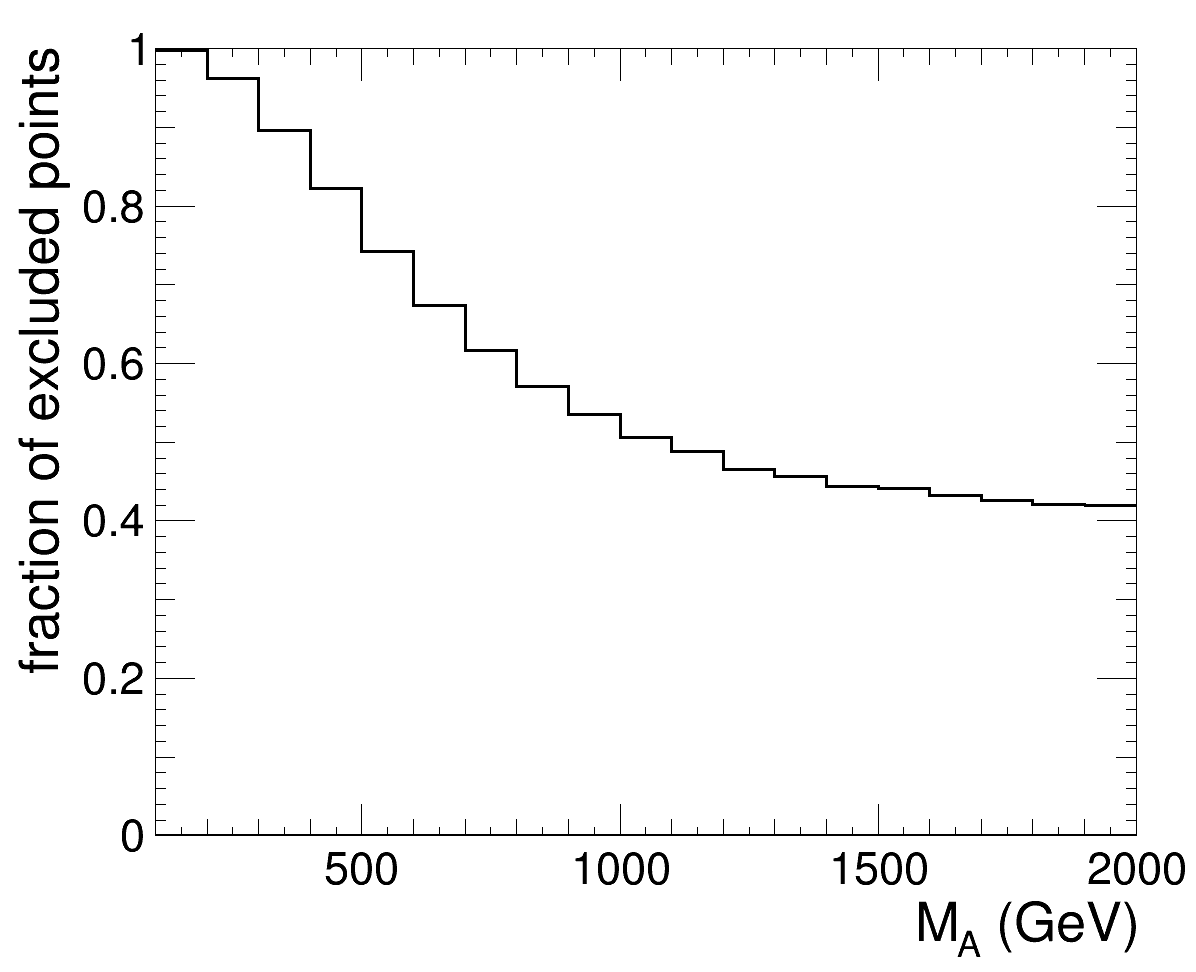}
\caption{Fraction of excluded points as a function of $M_A$ after applying the constraints from LUX on the dark matter direct detection spin-independent neutralino-proton scattering cross section.\label{fig:LUX2}}
\end{center}
\end{figure}

In Figs.~\ref{fig:LUX} and~\ref{fig:LUX2}, we present the constraints from LUX on $M_A$ and $\tan\beta$. As can be seen in the figures, important constraints are obtained by dark matter direct detection for large values of $\tan\beta$. We notice also that small $M_A < 200$ GeV are excluded by these searches independently of any other constraints, and that 90\% of the points with $M_A \lesssim 350$ GeV are ruled out.

\begin{figure}[t!]
\begin{center}
\includegraphics[width=7.cm]{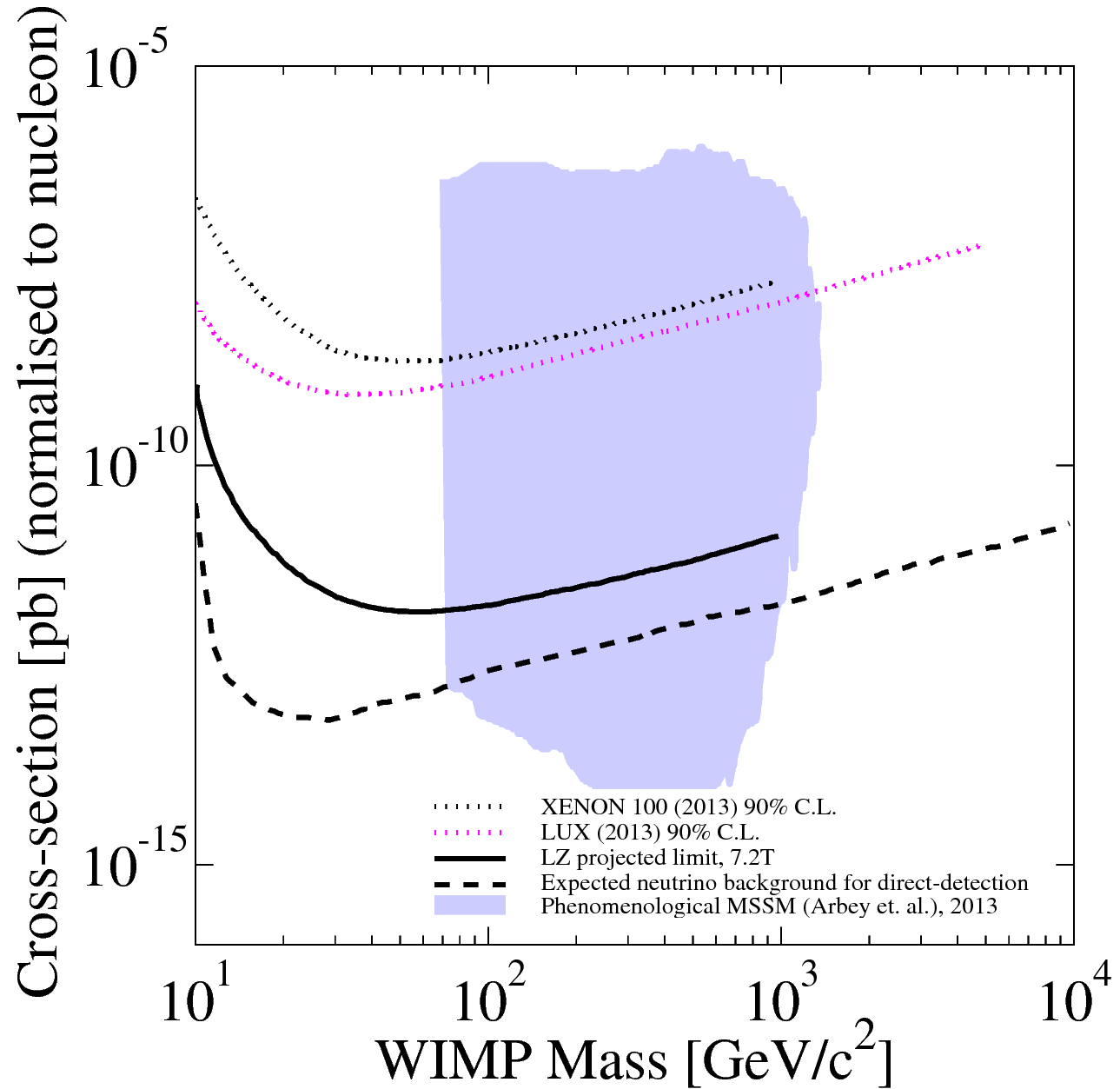}
\caption{Dark matter direct detection spin-independent scattering cross section of a neutralino with a nucleon as a function of the dark matter mass (from \cite{plotter}). The dotted red line corresponds to the current LUX limits. The solid black line is the projected limit of the future LZ experiment. The black dashed line corresponds to the neutrino background limit. The blue zone shows the spread of our pMSSM points in this plane.\label{fig:DD}}
\end{center}
\end{figure}%

In Fig.~\ref{fig:DD}, the spin-independent scattering cross section of a neutralino with matter is shown as a function of the dark matter mass, together with current and prospective limits of direct detection experiments~\cite{plotter}. The blue region represents the spread of our pMSSM points in this plane. 
The LUX limit excludes about 30\% of the model points. This fraction will become larger with the future experiment results becoming available, however, very small scattering cross sections can be obtained in the pMSSM so that even by reaching the neutrino background limit still a non-negligible fraction of the points would survive that could in a large extent be probed at the LHC high luminosity run or at future colliders.

\section{Conclusion}

In view of the negative results of the searches in channels with MET conducted by the ATLAS and CMS experiments at the end of the 7 and 8 TeV runs, it is clear that the simplest SUSY benchmarks are now ruled out and it is of great importance to reinterpret the results in more general scenarios with minimal theoretical assumptions. 
By studying the 19-parameter phenomenological MSSM, we have demonstrated that a wide parameter space of solutions with SUSY masses below 1 TeV, compatible with all the current data still exists.
Even at the end of the next LHC run, many of these solutions will not be tested. In such a situation, the only way to point to a specific scenario, or falsify the MSSM, would be to take advantage of interplay between different sectors. Regarding the LHC direct searches, in addition to direct SUSY search results, information from the Higgs sector namely from the mass and coupling measurements as well as information from other-than-SUSY searches such as mono-particle searches can be used to tightly constrain the parameter space. Moreover, indirect search results in particular from the flavour physics sector and rare decays, and also from the dark matter searches will allow us to deeply probe the MSSM parameter space. 
Such complementarities were discussed in this report. In particular, we showed that the monojet searches are sensitive to the regions with small mass splittings where the SUSY searches are weaker. Also the Higgs searches allow us to explore efficiently the ($M_A,\tan\beta$) plane which is also probed by both data from $B_s\to \mu^+\mu^-$ and from dark matter direct detection. These complementarities will be further exploited with the results of the next LHC runs, future dark matter detection experiments as well as the future Super-$B$ factory.

\section*{Acknowledgements}

FM and AA would like to thank the organisers for their invitation to the workshop and the participants for the many stimulating discussions. The results presented here have been achieved in collaboration with Marco Battaglia to whom the authors are grateful. The authors acknowledge partial support from the CNRS PEPS-PTI project ``DARKMONO''.



\begin{thebibliography}{00}

\bibitem{Arbey:2011un}
  A.~Arbey, M.~Battaglia and F.~Mahmoudi,
  Eur.\ Phys.\ J.\ C {\bf 72} (2012) 1847
  [arXiv:1110.3726 [hep-ph]].

\bibitem{Arbey:2011aa}
  A.~Arbey, M.~Battaglia and F.~Mahmoudi,
  Eur.\ Phys.\ J.\ C {\bf 72} (2012) 1906
  [arXiv:1112.3032 [hep-ph]].

\bibitem{Allanach:2001kg}
  B.~C.~Allanach,
  Comput.\ Phys.\ Commun.\  {\bf 143} (2002) 305
  [hep-ph/0104145].
  
\bibitem{Djouadi:1997yw}
  A.~Djouadi, J.~Kalinowski and M.~Spira,
  Comput.\ Phys.\ Commun.\  {\bf 108} (1998) 56
  [hep-ph/9704448].
 
\bibitem{Muhlleitner:2003vg}
  M.~Muhlleitner, A.~Djouadi and Y.~Mambrini,
  Comput.\ Phys.\ Commun.\  {\bf 168} (2005) 46
  [hep-ph/0311167].

\bibitem{Spira:1996if}
  M.~Spira,
  Nucl.\ Instrum.\ Meth.\ A {\bf 389} (1997) 357
  [hep-ph/9610350].

\bibitem{Harlander:2003ai}
  R.~V.~Harlander and W.~B.~Kilgore,
  Phys.\ Rev.\ D {\bf 68} (2003) 013001
  [hep-ph/0304035].

\bibitem{Spira:1997dg}
  M.~Spira,
  Fortsch.\ Phys.\  {\bf 46} (1998) 203
  [hep-ph/9705337].
  
\bibitem{Mahmoudi:2007vz}
  F.~Mahmoudi,
  Comput.\ Phys.\ Commun.\  {\bf 178} (2008) 745
  [arXiv:0710.2067 [hep-ph]];
  Comput.\ Phys.\ Commun.\  {\bf 180} (2009) 1579
  [arXiv:0808.3144 [hep-ph]].
  
\bibitem{Arbey:2009gu}
  A.~Arbey and F.~Mahmoudi,
  Comput.\ Phys.\ Commun.\  {\bf 181} (2010) 1277
  [arXiv:0906.0369 [hep-ph]].

\bibitem{Belanger:2008sj}
  G.~Belanger, F.~Boudjema, A.~Pukhov and A.~Semenov,
  Comput.\ Phys.\ Commun.\  {\bf 180} (2009) 747
  [arXiv:0803.2360 [hep-ph]].
  
\bibitem{Alwall:2011uj}
  J.~Alwall, M.~Herquet, F.~Maltoni, O.~Mattelaer and T.~Stelzer,
  JHEP {\bf 1106} (2011) 128
  [arXiv:1106.0522 [hep-ph]].

\bibitem{Sjostrand:2007gs}
  T.~Sjostrand, S.~Mrenna and P.~Z.~Skands,
  Comput.\ Phys.\ Commun.\  {\bf 178} (2008) 852
  [arXiv:0710.3820 [hep-ph]].

\bibitem{deFavereau:2013fsa}
  J.~de Favereau {\it et al.}  [DELPHES 3 Collaboration],
  JHEP {\bf 1402} (2014) 057
  [arXiv:1307.6346 [hep-ex]].

\bibitem{Ade:2013zuv}
  P.~A.~R.~Ade {\it et al.}  [Planck Collaboration],
  Astron.\ Astrophys.\  (2014)
  [arXiv:1303.5076 [astro-ph.CO]].

\bibitem{Arbey:2008kv}
  A.~Arbey and F.~Mahmoudi,
  Phys.\ Lett.\ B {\bf 669} (2008) 46
  [arXiv:0803.0741 [hep-ph]].

\bibitem{Arbey:2009gt}
  A.~Arbey and F.~Mahmoudi,
  JHEP {\bf 1005} (2010) 051
  [arXiv:0906.0368 [hep-ph]].

\bibitem{Arbey:2013aba}
  A.~Arbey, M.~Battaglia and F.~Mahmoudi,
  Phys.\ Rev.\ D {\bf 88} (2013) 095001
  [arXiv:1308.2153 [hep-ph]].
  
\bibitem{Arbey:2013iza}
  A.~Arbey, M.~Battaglia and F.~Mahmoudi,
  Phys.\ Rev.\ D {\bf 89} (2014) 077701
  [arXiv:1311.7641 [hep-ph]].

\bibitem{Djouadi:2005gj}
  A.~Djouadi,
  Phys.\ Rept.\  {\bf 459} (2008) 1
  [hep-ph/0503173].

\bibitem{Arbey:2011ab}
  A.~Arbey, M.~Battaglia, A.~Djouadi, F.~Mahmoudi and J.~Quevillon,
  Phys.\ Lett.\ B {\bf 708} (2012) 162
  [arXiv:1112.3028 [hep-ph]].
  
\bibitem{Arbey:2012dq}
  A.~Arbey, M.~Battaglia, A.~Djouadi and F.~Mahmoudi,
  JHEP {\bf 1209} (2012) 107
  [arXiv:1207.1348 [hep-ph]].

\bibitem{Arbey:2012bp}
  A.~Arbey, M.~Battaglia, A.~Djouadi and F.~Mahmoudi,
  Phys.\ Lett.\ B {\bf 720} (2013) 153
  [arXiv:1211.4004 [hep-ph]].
  
\bibitem{Arbey:2013jla}
  A.~Arbey, M.~Battaglia and F.~Mahmoudi,
  Phys.\ Rev.\ D {\bf 88} (2013) 1,  015007
  [arXiv:1303.7450 [hep-ph]].

\bibitem{CMSandLHCbCollaborations:2013pla}
  CMS and LHCb Collaborations,
  CMS-PAS-BPH-13-007, LHCb-CONF-2013-012, CERN-LHCb-CONF-2013-012.
  
\bibitem{Arbey:2012ax}
  A.~Arbey, M.~Battaglia, F.~Mahmoudi and D.~Martinez Santos,
  Phys.\ Rev.\ D {\bf 87} (2013) 035026
  [arXiv:1212.4887 [hep-ph]].

\bibitem{CMS-HIG-2012-050}
  CMS Collaboration,
  CMS-PAS-HIG-12-050.
  
\bibitem{Akerib:2013tjd}
  D.~S.~Akerib {\it et al.}  [LUX Collaboration],
  Phys.\ Rev.\ Lett.\  {\bf 112} (2014) 9,  091303
  [arXiv:1310.8214 [astro-ph.CO]].

\bibitem{Bernabei:2010mq}
  R.~Bernabei {\it et al.}  [DAMA and LIBRA Collaborations],
  Eur.\ Phys.\ J.\ C {\bf 67} (2010) 39
  [arXiv:1002.1028 [astro-ph.GA]].

\bibitem{Aalseth:2011wp}
  C.~E.~Aalseth, P.~S.~Barbeau, J.~Colaresi, J.~I.~Collar, J.~Diaz Leon, J.~E.~Fast, N.~Fields and T.~W.~Hossbach {\it et al.},
  Phys.\ Rev.\ Lett.\  {\bf 107} (2011) 141301
  [arXiv:1106.0650 [astro-ph.CO]].

\bibitem{Angloher:2011uu}
  G.~Angloher, M.~Bauer, I.~Bavykina, A.~Bento, C.~Bucci, C.~Ciemniak, G.~Deuter and F.~von Feilitzsch {\it et al.},
  Eur.\ Phys.\ J.\ C {\bf 72} (2012) 1971
  [arXiv:1109.0702 [astro-ph.CO]].

\bibitem{Ahmed:2010wy}
  Z.~Ahmed {\it et al.}  [CDMS-II Collaboration],
  Phys.\ Rev.\ Lett.\  {\bf 106} (2011) 131302
  [arXiv:1011.2482 [astro-ph.CO]].

\bibitem{Arbey:2012na}
  A.~Arbey, M.~Battaglia and F.~Mahmoudi,
  Eur.\ Phys.\ J.\ C {\bf 72} (2012) 2169
  [arXiv:1205.2557 [hep-ph]].
  
\bibitem{plotter}  
  J.~Filippini, R.~J.~Gaitskell, D.~Speller and G.~Wang, http://cedar.berkeley.edu/plotter/

\end{thebibliography}
\end{document}